\documentclass[aps,prb,twocolumn,superscriptaddress]{revtex4-1}

\usepackage{graphicx}
\usepackage{dcolumn}
\usepackage{bm}
\usepackage[normalem]{ulem}

\begin{document}

\title{Supersolidity in quantum films adsorbed on graphene and graphite
}  

\author{M.C. Gordillo}

\affiliation{Departamento de Sistemas F\'{\i}sicos, Qu\'{\i}micos 
y Naturales, Facultad de Ciencias Experimentales, Universidad Pablo de
Olavide, Carretera de Utrera, km 1. 41013 Sevilla, Spain}

\author{C. Cazorla}

\affiliation{
Institut de Ci$\grave{e}$ncia de Materials de Barcelona (ICMAB-CSIC), 
Campus UAB, 08193 Bellaterra, Spain 
}

\author{J. Boronat}

\affiliation{Departament de F\'{\i}sica i Enginyeria Nuclear, 
Universitat Polit\`ecnica de Catalunya, 
Campus Nord B4-B5, 08034 Barcelona, Spain}

\date{\today}

\begin{abstract}
Using quantum Monte Carlo we have studied the superfluid density of the
first layer of $^4$He and H$_2$ adsorbed on graphene and graphite. Our main
focus has been on the equilibrium ground state of the system, which corresponds to a
registered $\sqrt3 \times \sqrt3$ phase. The perfect solid phase of H$_2$ shows
no superfluid signal whereas $^4$He has a finite but small superfluid
fraction ($0.67$\%). The introduction of vacancies in the crystal makes the
superfluidity increase, showing values as large as $14$ \% in $^4$He without
destroying the spatial solid order. 
\end{abstract}


\maketitle

Supersolid state of matter is a fascinating possibility that has long attracted  
interest from both theoretical and experimental
viewpoints.~\cite{balibar,prokofev} The simultaneous existence of spatial lattice order and
off-diagonal long range order defining a supersolid is rather
counterintuitive and only a theoretical entelechy up to recent times. The
old theoretical ideas put forward by the pioneering works of Andreev and
Lifshitz~\cite{andreev} and Chester~\cite{chester} and Reatto~\cite{reatto}
 have revived dramatically since the
experimental findings of Kim and Chan~\cite{chan} in 2004 on the 
evidence of non-conventional moment of inertia in solid $^4$He below some
characteristic temperature around 100 mK. 
Whether these and some other similar experiments carried out by other teams are an 
unambiguous proof of the existence of supersolidity or not is still a matter of debate. 

At present, much less is known on possible supersolid scenarios in
two-dimensional (2D)~\cite{rossi} or quasi-two-dimensional (Q2D) solid $^4$He. 
Helium atoms, when adsorbed on graphite or graphene, 
arrange sequentially in stacking layers that can be considered as nearly 
two-dimensional systems.~\cite{cole} Interestingly, recent experiments 
carried out by Ny\'eki \textit{et al.}~\cite{nyeki} in the second layer of 
$^{4}$He adsorbed
on graphite point to the existence of a ($\sqrt{7} \times \sqrt{7}$)
commensurate solid
phase that exhibits superfluid fractions as large as 20\%. 
However, recent path integral Monte Carlo (PIMC) work,~\cite{bonin}  
does not seem to support the existence of this phase which was first predicted 
in simulations where the first layer of $^{4}$He atoms was considered as 
inert.~\cite{manou} Nevertheless,   
the possibility of having a supersolid in a 2D environment, supported 
by Ny\'eki \textit{et al.}'s observations,~\cite{nyeki} opens new and exciting 
avenues for the 
analysis of this new state of matter.

Recently, we have calculated the zero-temperature phase diagram of the first layer 
of $^4$He~\cite{carmen1} and H$_2$~\cite{carmen2} adsorbed on graphene and graphite using quantum Monte Carlo (QMC)
methods. Our results predict that the equilibrium ground state of both systems
is a $\sqrt{3} \times \sqrt{3}$ commensurate phase, conclusion
which, in the case of graphite, is in agreement with other low-temperature 
simulations and
experimental data.~\cite{dash1,dash2,dash3} The aim of our previous studies was basically the
determination of the energies of the different possible phases to draw the
phase diagram and not the study of off-diagonal long range order and/or
superfluidity.
To this end, we used a non-symmetric wave function for describing
the solid phases, an approach which obviously hinders any insight on the 
properties directly
related to their Bose-Einstein statistics but which guarantees accurate evaluation of the 
energies due to the low-rate interparticle exchange. 
In this work, we are mainly concerned with the possibility of
supersolidity in the first layer of $^4$He and H$_2$ adsorbed on graphene and graphite 
so that our methodology has been changed accordingly. 
Besides the characterization of the equilibrium ground-state phases, we
have also analyzed 
the influence of vacancies in the superfluid response and energy of these films 
since point defects are indeed observed during quantum layer nucleation on carbon 
surfaces.
It is worth mentioning that previous low-temperature attempts using PIMC have not found 
any signal of superfluidity in these layers~\cite{bonin,manou} although possible 
supersolidity induced by defects was not analyzed.

\begin{figure}[tbp]
\begin{center}
\includegraphics[width=0.85\linewidth]{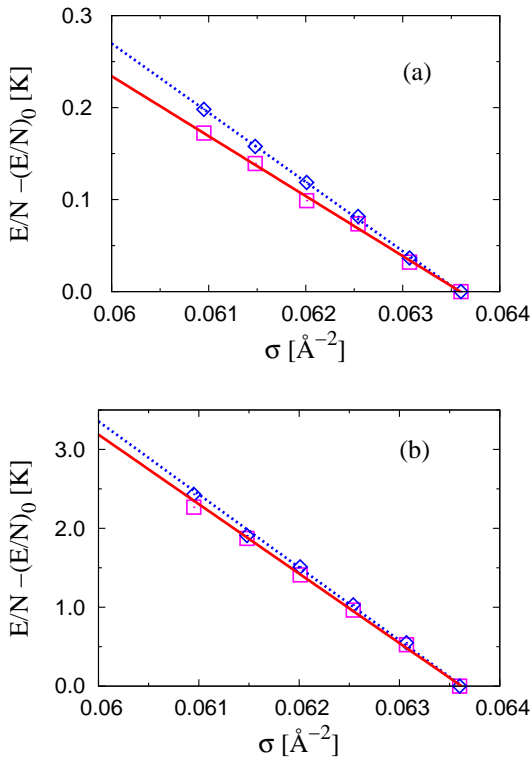}
\caption{(Color online) (a) Energy per $^4$He atom with different number of vacancies,
plotted as a function of the particle density; (b) Same as in (a) but
for H$_2$. In both cases the energy per particle of the commensurate phase with
no vacancies $(E/N)_0$ is subtracted. Solid line and squares (dotted line and
diamonds) stand for energies when the adsorbate is graphene (graphite).   
}
\end{center}
\label{fig:fig1}
\end{figure}

Since we are interested in possible ground-state supersolid phases of the
first $^4$He and H$_2$ layers adsorbed on graphene and graphite, we
use the diffusion Monte Carlo (DMC) method that, working at zero
temperature, solves stochastically the Schr\"odinger equation of the
$N$-particle system in an essentially exact way (within some statistical
uncertainties).~\cite{boro} Zero-temperature approaches are specially adequate for
these systems since the onset temperatures for supersolidity are, at least
in bulk solids, very small (50-100 mK)  thus making finite-temperature approaches
like PIMC of difficult use. The Hamiltonian of the system includes both
interparticle ($V(r)$) and particle-substrate ($U(r)$) interactions,
\begin{equation}
H=-\frac{\hbar^2}{2m} \sum_{i=1}^{N} {\bm \nabla}_i^2 + \sum_{i<j}^{N}
V(r_{ij}) + \sum_{i,J}^{N,N_s} U(r_{iJ}) \ ,
\label{hamilton}
\end{equation}
with capital indexes running on carbon substrate atoms and normal indexes on He
atoms or H$_2$ molecules. The $^4$He-$^4$He and H$_2$-H$_2$ interactions
are modeled by the accurate Aziz II~\cite{aziz} and
Silvera-Goldman~\cite{silvera} potentials,
respectively. The adsorbate surface potential is obtained by summing all the
pair atom(molecule)-adsorbate interactions modeled by Lennard-Jones
potentials. By summing up all the latter pair interactions, we introduce in
the description of the system the necessary corrugation to observe that a
commensurate solid phase is effectively preferred for being the ground
state.~\cite{carmen1,carmen2} 

As it is usual in DMC, we introduce a trial wave function for importance
sampling which improves the variance of the statistical estimation. Our
variational model contains basic information: it is zero at shorter
distances than the (hard-)core of the potential and becomes constant at large
distances (Jastrow wave function); it is symmetric under the exchange of particles 
(Bose-Einstein statistics), and localizes particles in preferred points
(sites) for solid phases. Explicitly,~\cite{claudi}
\begin{equation}
\Psi(\bm R) = \prod_{i<j}^{N} f(r_{ij}) \, \prod_{I=1}^{N_{\rm cr}}
\left[ \sum_{i=1}^{N} g(r_{Ii})  \right]  \ .
\label{psitrial}
\end{equation}   
In Eq. (\ref{psitrial}), ${\bm R} = \{ {\bm
r}_1,\ldots,{\bm r}_N \}$, $f(r)$ is a two-body Jastrow correlation factor, 
$g(r_{Ii})= \exp [- \alpha ({\bm r}_i -
{\bm r}_I)^2]$, and $N_{\rm cr}$ is the number of lattice sites of the
selected crystal structure. Model wave function (\ref{psitrial}) 
fulfills simultaneously spatial solid order and exchange-particle 
symmetry avoiding the numerically unworkable use of permanents on top
of the non-symmetric Nosanow-Jastrow wave function. The value of the 
variational parameters in $\Psi$ are the same as the ones reported in Refs.
\onlinecite{carmen1, carmen2}.

We have focused our attention in the $\sqrt{3} \times \sqrt{3}$
commensurate phase of $^{4}$He and H$_{2}$ since this is the equilibrium 
ground state structure and best candidate for exhibiting supersolid
behavior due to its quite low density. The simulation cell is a 
rectangle of fixed dimensions $44.27 \times 42.60$
\AA$^2$ where the number of sites of the registered phase is
$N_{\rm cr}=120$ and the number of particles is $115 \le N \le 120$
,i.e., we consider up to a maximum of five vacancies.
The dependence of the energy per particle with the number of vacancies,
reported as a function of the particle density, is shown in Fig. 1
for $^4$He and H$_2$ on top of graphene and graphite.
For comparison purposes, those curves have been displaced downwards an energy 
shift $(E/N)_0$ which corresponds to the energy of the perfect structures 
reported in Refs.~\onlinecite{carmen1,carmen2}.
For the present analysis it is enough  
to mention that the binding energy of H$_2$ is approximately $2.5$ times
larger than that of $^4$He and that in both systems the interactions with  
graphite are about 10~\% more attractive than with graphene.

As shown in Fig~1, the binding energies per particle increase
 systematically with the number of vacancies so  
the equilibrium ground state always corresponds to the perfect commensurate lattice. 
The variation of the activation vacancy energy is linear in both cases   
and slightly larger in graphite where the adsorption 
energies are invariably $\sim 10$ \% larger than in graphene. 
Comparing the results obtained
for $^4$He and H$_2$, one can see that the energy difference $E/N-(E/N)_0$
is roughly 10 times larger for hydrogen  however, as the absolute values for
both solids are so different, it is better to make a comparison in relative
terms:
the energy cost of creating 5 vacancies in $^4$He is 0.15 \% of the ground-state
energy while in H$_{2}$ is 0.43~\%. When
the number of vacancies increases, and therefore the particle density
decreases, the equation of state of the solid with vacancies approaches
the equation of state of the metastable liquid phase. This analysis is
specially interesting when graphene is the substrate since the difference
in binding
energy at the equilibrium point of the liquid and the one at the
commensurate solid phase is very tiny, nearly 4 times smaller than in
graphite. For both $^4$He and H$_2$, all the energies shown in Fig.~1 
are below the ones calculated for the corresponding liquid phases at the same densities. 
If the linear behavior observed in the figure is extrapolated to smaller densities, 
one can see that in $^4$He the intersection 
with the liquid equation of state is produced at density $\sigma_{\rm c}=0.058$
\AA$^{-2}$, which would correspond to 10 vacancies in the simulation box.
In this case, $\sigma_{\rm c}$ is larger than the equilibrium density of
the liquid $\sigma_0=0.044$ \AA$^{-2}$ thus the crossing point will
appear at finite pressure. In contrast, the intersection with the liquid 
equation of state in hydrogen appears at density $\sigma_{\rm c}=0.052$ 
\AA$^{-2}$, which is smaller than the equilibrium point  
$\sigma_0=0.059$ \AA$^{-2}$ and thus corresponds to the metastable negative 
pressure regime.

\begin{figure}[tbp]
\begin{center}
\includegraphics[width=0.85\linewidth]{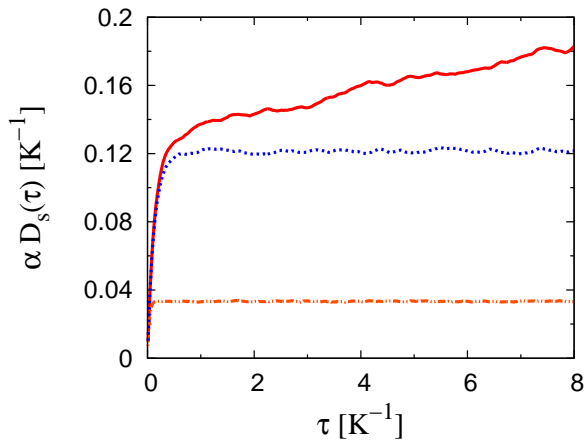}
\caption{(Color online) DMC estimation of the superfluid density. Solid,
dotted, and dashed lines stand for commensurate $^4$He, 
incommensurate $^4$He, and commensurate H$_2$ phases on top of graphene, respectively. 
}
\end{center}
\label{fig:fig2}
\end{figure}

A discussion on possible supersolid phases of helium and hydrogen on top of
graphene or graphite requires the estimation of the superfluid density
fraction of the different solid phases that appear in their respective  
phase diagrams. In quantum Monte Carlo, the superfluid fraction can be
computed by sampling the winding number~\cite{ceper}      
\begin{equation}
\bm{W} = \sum_{i=1}^{N} \int_0^{\beta} d\tau \left( \frac{d
\bm{r}_i(\tau)}{d \tau} \right) \ ,
\label{wind}
\end{equation}     
with $\tau$ the imaginary time and $\beta=T^{-1}$. In the limit of zero
temperature $\beta \to \infty$, the winding number (\ref{wind}) turns to
the diffusion coefficient of the center of mass of the $N$ particles
($\bm{R}_{CM}$) in the
simulation box with periodic boundary conditions,~\cite{zhang}
\begin{equation}
\frac{\rho_{\rm s}}{\rho} = \lim_{\tau \to \infty} \alpha \left(
\frac{D_{\rm s}(\tau)}{\tau} \right) \ ,
\label{wind2}
\end{equation}  
where $\alpha=N/2 d D_0$, with $d$ the number of dimensions ($d=2$ in the
present case),
$D_0=\hbar^2/2m$, and $D_{\rm
s}(\tau)=\langle (\bm{R}_{CM}(\tau) - \bm{R}_{CM}(0))^2 \rangle$. The
diffusion coefficient $D_{\rm s}(\tau)$ can be calculated using the DMC
method, in which the imaginary time evolves in a continuous way, and it can
be proved that in the asymptotic regime this estimator is unbiased with
respect to the particular choice of the trial wave function used for
importance sampling (within a specified physical phase). According to the
typical statistical noise in DMC simulations, the resolution of this
estimator is much lower ($\sim 1 \times 10^{-5}$) than in PIMC, where
superfluid signals below $\sim 1 \times 10^{-2} $ are extremely difficult
to be measured. 

\begin{figure}[tbp]
\begin{center}
\includegraphics[width=0.85\linewidth]{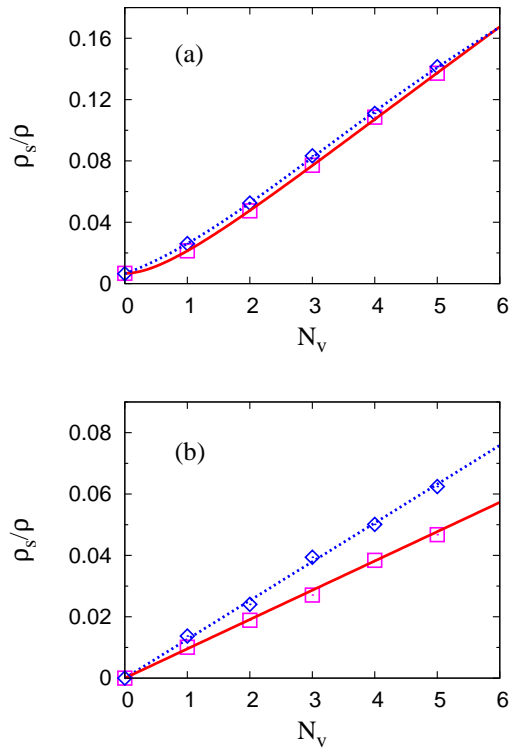}
\caption{(Color online) (a) Superfluid fraction of the $^4$He commensurate
solid phase  with different number of vacancies $N_v$; (b) Same as in (a) but
for H$_2$.  Solid line and squares (dotted line and
diamonds) stand for superfluid fractions when the adsorbate is graphene (graphite).   
}
\end{center}
\label{fig:fig3}
\end{figure}

In Fig. 2, results for  $\alpha D_{\rm s}(\tau)$ as a function of
the imaginary time $\tau$ are shown. 
As obvious from its definition
(\ref{wind2}), a finite superfluid fraction appears as a finite slope in
the long-time behavior of the diffusion coefficient $D_{\rm s}(\tau)$,
their particular values being not relevant and fairly dependent of the kind of
system under study. 
Our results for the perfect (no vacancies) solid
phases plotted in Fig. 2 show different behaviors depending on the
system and solid phases considered. Diffusion coefficient results for long
time $\tau$  
obtained for the $\sqrt{3} \times \sqrt{3}$
commensurate phase of $^4$He on top of graphene (and also in graphite)
show a small but clear slope ($\rho_s/\rho= 0.0067(1)$) that  contrasts with the 
null $\tau$-variation observed in a incommensurate phase of 
density $0.0999$ \AA$^{-2}$. 
Interestingly, simulations performed in the commensurate phase of H$_2$ 
indicate zero superfluid fraction thus areal density must be ruled out
as the only cause behind supersolidity in quantum films. 
Moreover, as it was shown in the DMC calculation of
2D and quasi-2D $^4$He in Ref. \onlinecite{claudi2}, the superfluid fraction
 of a purely two-dimensional crystal is zero even at densities below the               
$\sqrt{3} \times \sqrt{3}$ phase, and  finite superfluidity emerges  only with
the opening of a transverse direction that particles can explore. 
According to these previous results, the zero signal observed in H$_2$ can be
explained in terms of transverse motion frustration  resulting from  
intense molecular binding to carbon surfaces.

\begin{figure}[tbp]
\begin{center}
\includegraphics[width=0.9\linewidth]{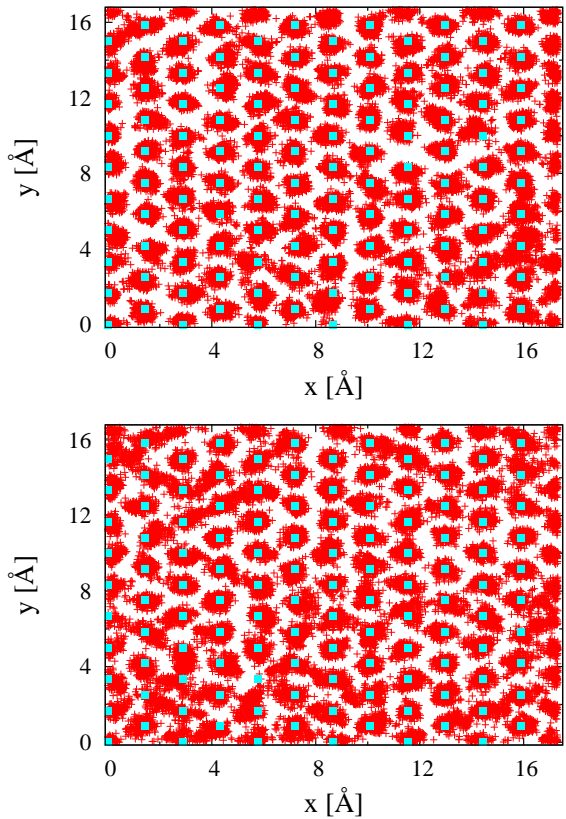}
\caption{(Color online) \textit{Top:} Snapshot of the probability density
(crosses) of commensurate solid $^4$He on top of graphene.  
 \textit{Bottom:} Same as in the top panel but with five less particles in
 the simulation box. In both figures, squares are the sites of the
 perfect crystal. 
}
\end{center}
\label{fig:fig4}
\end{figure}

One of the aims of this work has been the study of the influence of the
number of vacancies in the superfluid fraction of quantum solid layers. 
To this end, we have calculated $\rho_s/\rho$ for the commensurate phase of $^4$He
and H$_2$ on top of graphene and graphite with up to five ($N_v$) vacancies within a
total of 120 possible sites of our simulation cell. The results, shown in
Fig. 3, show a linear increase of the superfluid fraction with
$N_v$ for both helium and  hydrogen.
The increase of superfluidity with $N_v$ is
significantly larger for $^4$He; for instance, in the $N_v=5$
case  $\rho_s/\rho$ amounts to $0.14$ and  $0.047$ for $^4$He and H$_2$, 
respectively. The effect of the
substrate on the superfluid fraction is smaller in helium than in hydrogen
but in both cases $\rho_s/\rho$ is always larger in graphite than in 
graphene. The difference is very small in $^4$He but significant in
H$_2$. A possible explanation of the increase of superfluidity in graphite
relies on the slightly reduced motion in the transverse direction due to 
stronger attraction with respect to graphene. This increase in
confinement makes more effective the motion of the center of mass 
of the system in the $x-y$ plane of the simulation box, where large
particle-permutational rings can be created, and consequently $\rho_s/\rho$ 
increases~(\ref{wind2}).

We have verified that both in hydrogen and helium the periodic spatial
order characteristic of the solid phase is conserved up to 5 vacancies
in 120 possible sites. This is shown in Fig. 4, where
characteristic snapshots of DMC simulations are shown for the perfect 
and defective $N_{v} = 5$ film phases ($^4$He). These snapshots correspond 
to the probability density that in DMC is represented by a collection of walkers, 
each one of $3N$ coordinates, that evolves in 
imaginary time according to the Schr\"odinger
equation. As one can see, even in the case of the perfect crystal there is
a finite probability of visiting the intersite space, which can be
interpreted as the finite exchange probability leading to non-zero
superfluid signal (a quantitative estimation of probability exchange would
require of specific methods beyond the scope of the present
work~\cite{ceperrmp}). 
When vacancies are present in the system 
the paths connecting different sites become more populated, i.e.,
the superfluidity increases, but solid order remains preserved. It is
worth noticing that in the snapshot of the layer with vacancies one can not
allocate the vacancies implying that they have become indistinguishable.
Same snapshots for H$_2$ show that in the perfect crystal the intersite
occupation is zero (zero superfluidity) and that the vacancies are identified more
easily than in helium.

Summarizing, using the DMC method we have studied the supersolidity of the
first layer of $^4$He and H$_2$ adsorbed on graphene and graphite in the
limit of zero temperature. The  $^4$He $\sqrt{3} \times \sqrt{3}$ commensurate
phase shows a small but finite superfluid signal ($0.67$ \%) whereas the
H$_2$ one does not within our resolution limit ($1 \times 10^{-5}$). 
When vacancies are present in the system, the superfluid
fraction increases with the concentration of defects; this effect 
is larger in helium where we have obtained values as large as 
$\rho_s/\rho=14$ \%. As the presence of point defects in quantum 
layers is plausible due to imperfections in the adsorbent surfaces, 
further experiments on quasi-2D systems can lead to the emergence of new 
and tunable supersolid scenarios.~\cite{nyeki,shiba}

Authors would like to thank Jan Ny\'eki for helpful discussions.
We acknowledge partial financial support from the 
Junta de Andaluc\'{\i}a group PAI-205, DGI (Spain) Grant No.
 FIS2008-04403 and Generalitat de Catalunya Grant No. 2009SGR-1003.



\begin{thebibliography}{99}

\bibitem{balibar} S. Balibar and F. Caupin, J. Phys.: Condens. Matter
\textbf{20}, 173201 (2008).

\bibitem{prokofev} N. Prokof'ev, Adv. Phys. \textbf{56}, 381 (2007).

\bibitem{andreev} A. F. Andreev and I. M. Lifshitz, Sov. Phys. JETP
\textbf{29}, 1107 (1969).

\bibitem{chester} G. V. Chester, Phys. Rev. A \textbf{2}, 256 (1970).

\bibitem{reatto} L. Reatto, Phys. Rev.  \textbf{183}, 334 (1969).
  
\bibitem{chan} E. Kim and M. H. W. Chan, Science \textbf{305}, 1941 (2004);
Nature \textbf{427}, 225 (2004).

\bibitem{rossi} M. Rossi, E. Vitali, D. E. Galli, and L. Reatto,  J. Phys.: Condens. Matter 
\textbf{22}, 145401 (2010).

\bibitem{cole} L. W. Bruch, M. W. Cole, and E. Zaremba, \textit{Physical
Adsorption: Forces and Phenomena} (Oxford Science Publishers, Oxford,
1997).

\bibitem{nyeki} Communication presented by J. Ny\'eki \textit{et al.} to the workshop 
Supersolids 2009 held in Banff (Canada).

\bibitem{bonin} P. Corboz, M. Boninsegni, L. Pollet, and M. Troyer, Phys.
Rev. B \textbf{78}, 245414 (2008).

\bibitem{manou} M. Pierce and E. Manousakis, Phys. Rev. Lett. \textbf{81},
156 (1998).

\bibitem{carmen1} M. C. Gordillo and J. Boronat, Phys. Rev. Lett.
\textbf{102}, 085303 (2009).

\bibitem{carmen2} M. C. Gordillo and J. Boronat, Phys. Rev. B \textbf{81},
155435 (2010).

\bibitem{dash1} M. Bretz and J. G. Dash, Phys. Rev. Lett. \textbf{27}, 647
(1971).

\bibitem{dash2} M. Bretz, J. G. Dash, D. C. Hickernell, E. O. McLean, and
O. E. Vilches, Phys. Rev. A \textbf{8}, 1589 (1973).

\bibitem{dash3} J. G. Dash, Phys. Rev. B \textbf{15}, 3136 (1977).

\bibitem{boro} J. Boronat and J. Casulleras, Phys. Rev. B \textbf{49}, 8920
(1994).

\bibitem{aziz} R. A. Aziz, F. R. W. McCourt, and C. C. K. Wong., Mol. Phys.
\textbf{61}, 1487 (1987). 


\bibitem{silvera} I. F. Silvera and V. V. Goldman, J. Chem. Phys. \textbf{69}, 4209
(1978).


\bibitem{claudi} C. Cazorla, G. E. Astrakharchik, J. Casulleras, and J.
Boronat, New J. Phys \textbf{11}, 013047 (2009).

\bibitem{ceper} D. M. Ceperley, Rev. Mod. Phys. \textbf{67}, 279 (1995).

\bibitem{zhang} S. Zhang, N. Kawashima, J. Carlson, and J. E. Gubernatis,
Phys. Rev. Lett. \textbf{74}, 1500 (1995).

\bibitem{claudi2} C. Cazorla, G. E. Astrakharchik, J. Casulleras, and J.
Boronat, J. Phys.: Condens. Matter  \textbf{22}, 165402 (2010).

\bibitem{ceperrmp} D. M. Ceperley, Rev. Mod. Phys. \textbf{67}, 279 (1995).

\bibitem{shiba} Y. Shibayama, H. Fukuyama, and K Shirahama, J. Phys.: Conf.
Ser. \textbf{150}, 032096 (2009).


\end{thebibliography}
\end{document}